\documentclass[sigconf, authorversion]{acmart}
\AtBeginDocument{%
}

\copyrightyear{2026}
\acmYear{2026}
\setcopyright{cc}
\setcctype{by}
\acmConference[RecSys '26]{20th ACM Conference on Recommender Systems}{September 27--October 02, 2026}{Minneapolis, MN, USA}
\acmBooktitle{20th ACM Conference on Recommender Systems (RecSys '26), September 27--October 02, 2026, Minneapolis, MN, USA}
\acmDOI{10.1145/3773078.3831848}
\acmISBN{979-8-4007-2284-4/2026/09}

\begin{document}

\title{Heterogeneous Ranking in Industrial-Scale Recommender Systems: A Case Study}


\author{Di Bai}
\affiliation{%
\institution{Google LLC}
\city{Mountain View}
\state{CA}
\country{USA}}
\email{vivianbai@google.com}
\orcid{0009-0009-1813-4844}

\author{Jintao Liu}
\affiliation{%
\institution{Google LLC}
\city{Mountain View}
\state{CA}
\country{USA}}
\email{liujintao@google.com}
\orcid{0009-0005-1325-0142}

\author{Zhenwei Tang}
\affiliation{%
\institution{Google LLC}
\city{Toronto}
\state{ON}
\country{Canada}}
\email{lilvjosephtang@google.com}
\orcid{0000-0002-8742-9146}

\author{Peifan Wu}
\affiliation{%
\institution{Google LLC}
\city{New York}
\state{NY}
\country{USA}}
\email{peifanwu@google.com}
\orcid{0009-0006-8684-415X}

\author{Nada Al-Thawr}
\affiliation{%
\institution{Google LLC}
\city{San Francisco}
\state{CA}
\country{USA}}
\email{nalthawr@google.com}
\orcid{0009-0006-7396-630X}

\author{Luoshu Wang}
\affiliation{%
\institution{Google LLC}
\city{Mountain View}
\state{CA}
\country{USA}}
\email{luoshu@google.com}
\orcid{0009-0009-6681-545X}

\renewcommand{\shortauthors}{Bai et al.}

\begin{abstract}
Heterogeneous recommendation feeds present complex challenges that extend beyond those found in highly homogeneous environments (e.g., music-only or video-only closed-ecosystem platforms). In Google Discover, a unified feed integrates diverse content sourced from the decentralized open web, including web articles, long-form and short-form videos, user-generated content (UGC), and beyond. Different content types exhibit distinct feature densities and user interaction patterns. Building a unified ranking model that sustains high performance across such heterogeneity, while avoiding negative transfer or majority bias, remains a significant industrial challenge.

This paper presents an end-to-end case study on the industrial-scale multi-task ranking of heterogeneous feeds, grounded in real-world deployment. We introduce HA-MoE, a heterogeneity-adaptive multi-gated mixture-of-experts architecture that incorporates explicit heterogeneity context into both gating networks and expert representations. This approach enables effective specialization without significantly increasing operational overhead. To support reliable deployment, we introduce LENS, a lightweight observability framework that provides interpretable diagnostics of expert specialization and tracks this functional heterogeneity across continuous retraining. We evaluate our method using Dual-Level AUC (DL-AUC), a heterogeneity-aware evaluation metric that combines global ranking performance with cross-segment ranking correctness. Offline evaluations on a large-scale industrial dataset demonstrate consistent improvements over baseline models. Furthermore, online A/B testing confirms gains in feed activity and exploration metrics. Together, offline and online results validate the effectiveness of our approach for managing heterogeneity in industrial-scale recommender systems.
\end{abstract}



\maketitle

\section{Introduction}
\begin{figure}[htbp]
\centering
\includegraphics[width=\linewidth]{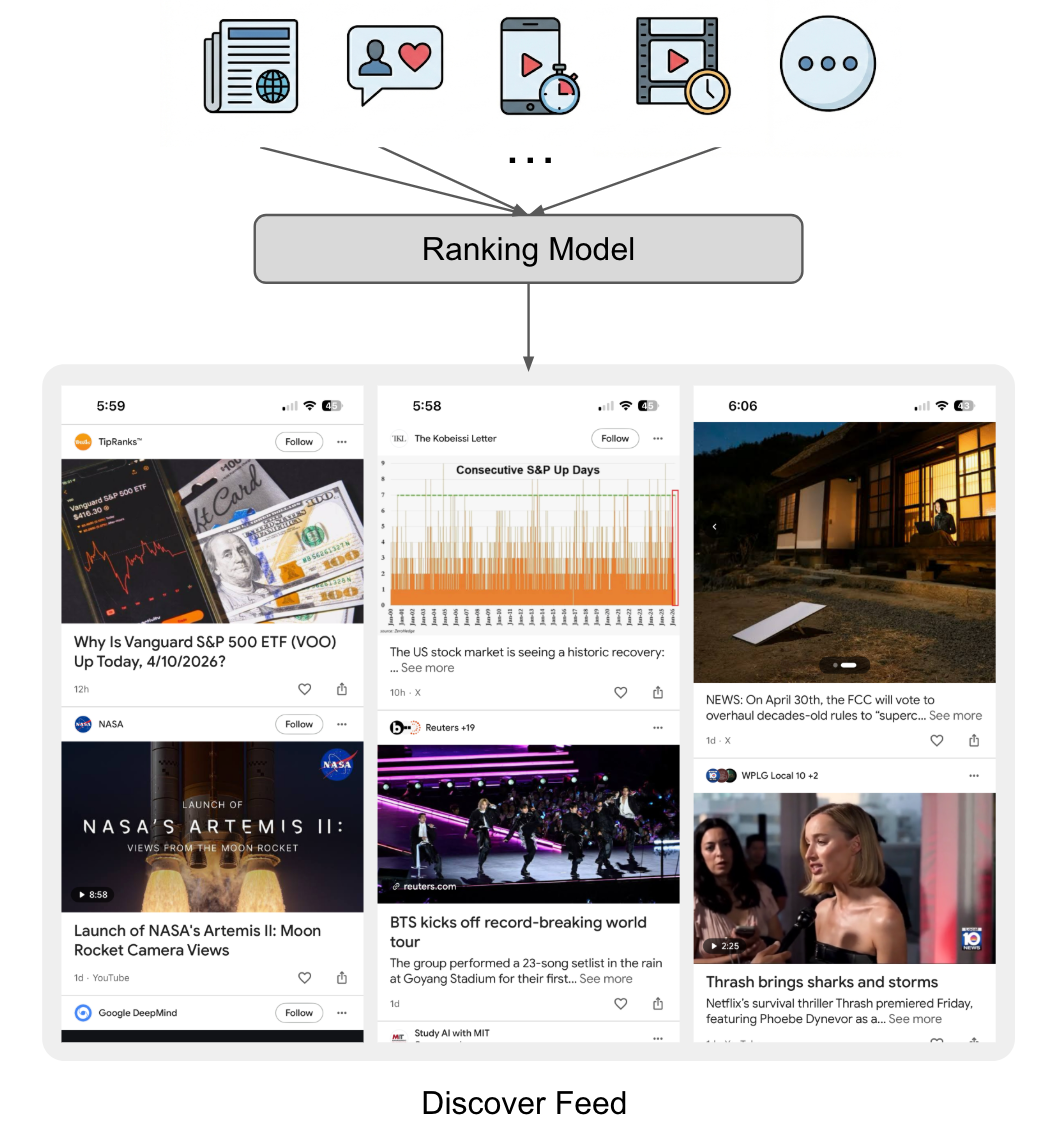}
\Description{An overview illustration of Google Discover, showing how heterogeneous content sourced from the open web is scored by a single unified ranking model and surfaced as a personalized feed.}
\caption{\textbf{Overview of Google Discover.} The system constructs a personalized feed by evaluating heterogeneous content from the open web through a unified ranking model.}
\label{fig:discover_overview}
\end{figure}

Many recommender systems are built for self-contained platforms, such as dedicated short-video or music streaming applications. In these relatively homogeneous, closed ecosystems, items share similar modalities and interaction patterns, and upstream signals are standardized across the platform. Even when feeds blend heterogeneous content---such as organic and ad cards---they typically rely on separate ranking models that are merged only during a late-stage re-ranking pass~\cite{pei2019personalized, zhao2021dear}. Google Discover, by contrast, uses a single ranking model to construct a unified, personalized feed from the highly heterogeneous open web (Figure~\ref{fig:discover_overview}).

The system must jointly evaluate a wide spectrum of content types, including standard web articles, long- and short-form videos, and user-generated content (UGC) posts, alongside specialized formats like sports match cards and generative AI summaries. This content heterogeneity presents a unique challenge that spans two dimensions:

\begin{itemize}
\item \textit{Data Heterogeneity:} Content sourced from the open web lacks uniform features, causing inherent variations in feature densities and metadata structures, as well as asymmetric signal availability.
\item \textit{Interaction Heterogeneity:} Engagement patterns and user intents vary. For instance, while likes and dismissals apply broadly, articles primarily drive clicks, whereas UGC posts and sports match cards drive long dwell time.
\end{itemize}

This inherent content heterogeneity introduces several practical challenges for unified ranking. First, distribution mismatch occurs because engagement distributions vary drastically across content types. A model optimized primarily for global click-through rate may disproportionately favor low-quality articles (e.g., click-bait) over high-value videos and UGC, leading to a sub-optimal user experience. Second, multi-task ranking systems that jointly optimize conflicting objectives, such as maximizing clicks while minimizing dismissals, often suffer from the seesaw phenomenon, where improvements in one objective negatively affect others due to negative transfer. Finally, shared-representation architectures carry the risk of minority-type collapse by favoring dominant content types with richer features or higher traffic. 

The current production baseline relies on a shared Multi-Layer Perceptron (MLP) backbone with 11 task-specific prediction heads. These heads predict a mix of positive and negative user interactions, covering both universal engagements (e.g., predicted click-through rate, dismissal rate) and format-specific actions. While efficient, this architecture forces a single latent representation to serve all content types. Public datasets~\cite{dror2012yahoo, chen2023reasoner, yuan2022tenrec, gao2022kuairec} lack the content heterogeneity required to study such dynamics. We therefore rely exclusively on our in-house large-scale industrial dataset.

To address these challenges, we present a production case study on deploying unified multi-task ranking models over heterogeneous feeds at industrial scale. We describe \textbf{HA-MoE}, a heterogeneity-adaptive multi-gated mixture-of-experts architecture designed to improve specialization while respecting constraints on model size, training speed, and serving latency. We also introduce \textbf{LENS}, an analytical framework that ``unboxes'' internal model dynamics. LENS empowers engineers to interpret the model's internal division of labor and quantitatively track functional task specialization across continuous training cycles. Finally, we evaluate our method using \textbf{Dual-Level AUC (DL-AUC)}, which measures ranking performance at two levels by balancing global utility and cross-segment correctness. Together, these components form an end-to-end workflow covering model design, evaluation, and deployment guardrails for reliable heterogeneous ranking systems in production.

\section{Related Work}
\subsection{Multi-Task and Multi-Domain Recommendation}
Early multi-task learning in recommender systems mainly relied on shared-bottom Multi-Layer Perceptrons (MLPs) to jointly optimize diverse user engagement signals \cite{caruana1997multitask, covington2016deep}. To mitigate the inherent negative transfer between competing objectives in these shared representations, foundational architectures like Multi-gate Mixture-of-Experts (MMoE) \cite{ma2018modeling} and Progressive Layered Extraction (PLE) \cite{tang2020progressive} introduced task-specific gating and decoupled experts. Recent advancements have further refined these frameworks by dynamically balancing parameter updates for conflicting tasks \cite{yuan2024parameter}.

Despite these architectural designs, handling highly heterogeneous content within a unified model remains a distinct challenge. Consequently, many industrial feeds bypass unified modeling entirely, merging domain-specific ranker outputs downstream via late-stage heuristics, reserved slots, or auction-based bidding mechanisms \cite{zhao2021dear}. To bridge this gap, Multi-Domain Recommendation (MDR) architectures such as STAR \cite{sheng2021one}, PEPNet \cite{chang2023pepnet} and M\textsuperscript{3}oE \cite{zhang2024m3oe} have advanced unified serving by incorporating conditional parameter modulation. These methods, including Feature-wise Linear Modulation (FiLM) \cite{perez2018film} and its adaptations for structured data \cite{brockschmidt2020gnn}, utilize external signals to adapt internal activations or weights, thereby aligning disparate distributions without excessive parameter scaling. Furthermore, recent weight-ensembling approaches \cite{tang2024merging} have been employed within MoE layers to mitigate parameter interference. While these modulation techniques facilitate cross-domain alignment, recent work by Zhao et al.~\cite{zhao2024retrievable} has also begun to address feature-level heterogeneity by identifying domain-sensitive features and utilizing a retrievable memory mechanism to explicitly capture domain-specific distinctions. 
Nevertheless, effectively handling this heterogeneity within the constraints of industrial-scale, production-ready ranking systems remains a critical open challenge.


\subsection{Heterogeneous Ranking Systems}


In the evolving landscape of industrial recommender systems, the focus has shifted from optimizing single-format feeds to managing heterogeneous environments where items, objectives, and sources vary drastically. Industrial applications of heterogeneous ranking fall into several distinct categories:

\textbf{Intra-platform Content Heterogeneity}: Feeds like TikTok blend diverse content formats (e.g., short videos, long-form videos, and image-text posts). The primary challenge lies in the disparity of information density and user interaction patterns. For instance, video consumption is measured by watch time, while posts rely on engagement like comments. Recent advancements explore unified representation learning to bridge disparate feature spaces \cite{lu2022deep}, deploy heterogeneous graph neural networks for ad-keyword matching \cite{liu2021heterogeneous}, and utilize reinforcement learning to optimize recommendations across diverse item channels \cite{xie2021hierarchical}, preventing ``majority bias'' where dominant formats overshadow high-quality niche content.

\textbf{Cross-business Objective Heterogeneity}: Most recommender systems come along with commercial needs. Within the feed shown to users, platforms must jointly rank organic content based on user engagement and advertisements with consideration for monetization. While these items may share a similar appearance, their optimization goals are often different and sometimes conflicting. As discussed in \cite{li2024ad}, the ``seesaw phenomenon'' is a common bottleneck, where commercial gains regress user retention. Usually, these systems use a two-stage mechanism rather than fully unified models to balance the conflicting objectives from different businesses \cite{pei2019personalized, liu2018personalizing}, where the first stage deals with the intra-ranking separately within each specific type and the second stage merges the ordered list for final exposure to users. 

Unlike closed-ecosystem platforms that benefit from standard, unified signal processing, platforms such as Google Discover navigate a unique structural complexity by aggregating content from the decentralized open web. This diverse ecosystem precludes strict upstream feature standardization. Consequently, the system must process inherently non-uniform feature spaces, leading to variations in data density and asymmetric signal availability. Furthermore, user engagement patterns naturally shift across different content modalities, necessitating the alignment of multiple optimization objectives directly within the ranking model. This inherent heterogeneity reflects the core paradigm of web-scale retrieval and ranking: aggregating candidate items across disparate sources and surfacing them within a unified feed \cite{wang2016beyond}. Recent literature has explored strategies to address these challenges, primarily through unified feature mapping for multi-source integration \cite{lu2022deep} and the incorporation of side-information to mitigate issues related to sparse or inconsistent feature sets \cite{liu2019recommender}.
However, translating these academic strategies into robust and deployable systems that satisfy the production constraints of industrial-scale feeds remains an ongoing challenge.

\section{Methodology}

In this section, we detail our approach to robust and scalable unified ranking for highly heterogeneous feeds. Our methodology is guided by strict industrial production constraints, including training speed, serving latency, and operational stability. We first formalize the multi-task ranking problem and the integration of explicit heterogeneity signals (Section \ref{sec:problem_formulation}). Next, we introduce a practical adaptation of the Mixture-of-Experts framework that leverages these signals for adaptive gating and feature modulation without massive parameter expansion (Section \ref{sec:ha_moe_architecture}). Finally, to ensure reliable continuous deployment, we present a production monitoring framework designed to interpret expert specialization and safeguard structural stability across routine model refreshes (Section \ref{sec:lens_framework}).

\subsection{Problem Formulation}
\label{sec:problem_formulation}

The primary objective is to construct a unified, personalized feed by evaluating a diverse corpus of heterogeneous items $\mathcal{I}$ for a set of users $\mathcal{U}$.
Each user-item interaction is evaluated based on standard dense features and explicit heterogeneity signals.
Let $x \in \mathbb{R}^d$ denote the dense feature representations of the input, and $h \in \mathbb{R}^{d_h}$ represent the encoded explicit heterogeneity signals.
These signals directly encode the structural context of the content, spanning categorical attributes like \texttt{content\_type} (e.g., web articles vs. UGC posts) and a broad suite of contextual metadata, such as followed-creator status, shoppable content, and AI-enhanced cards.

To capture diverse user intents, the system must jointly predict $K=11$ distinct user interaction tasks, denoted by the set $\mathcal{T} = \{1, 2, \dots, K\}$. 
These tasks cover universal engagements (e.g., predicted click-through rate, dismissal rate) and format-specific actions.
We group these outputs into two macro-categories: \textit{pInterest} (positive engagements) and \textit{pDisinterest} (negative engagements).
For a user $u \in \mathcal{U}$ and item $i \in \mathcal{I}$, let $y_{t} \in \{0, 1\}$ denote the ground truth binary interaction label for task $t \in \mathcal{T}$. The unified ranking model learns a mapping function to predict the probabilities $\mathbf{p} \in [0, 1]^K$.

\subsection{Heterogeneous Ranking via HA-MoE}
\label{sec:ha_moe_architecture}

\begin{figure}[t!]
\centering
\includegraphics[width=\linewidth]{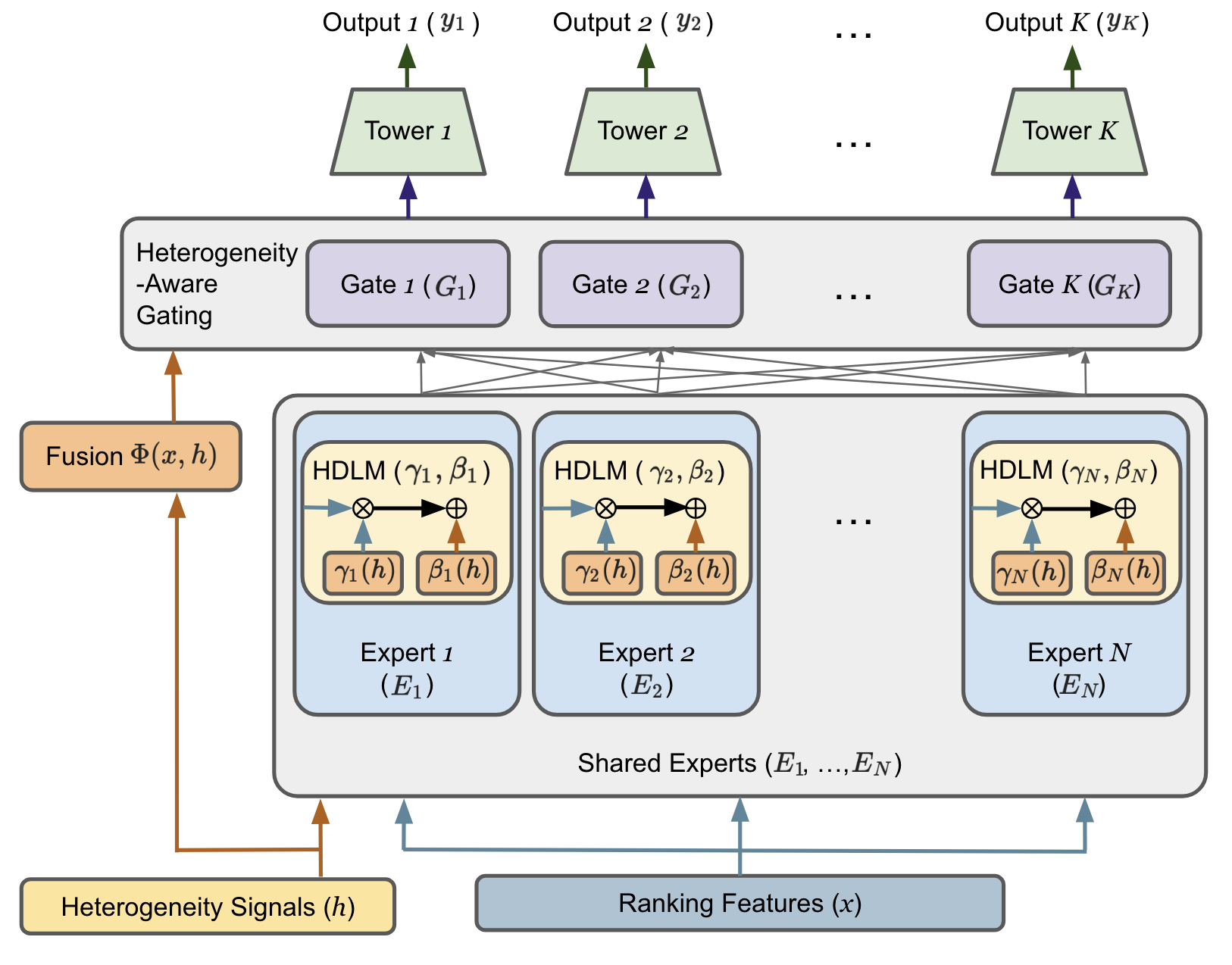}
\Description{Block diagram of HA-MoE. The explicit heterogeneity signals feed both the per-task gating networks and an affine modulation applied to each shared expert. For each of the K tasks, the gate-weighted sum of modulated experts is passed through a task-specific tower to produce a prediction.}
\caption{Overview of HA-MoE. Explicit heterogeneity signals are injected into both the multi-task gating network and the expert representations to adaptively handle disparate content distributions without massive parameter expansion.}
\label{fig:ha_moe}
\end{figure}

Deploying unified ranking architectures at industrial scale imposes strict operational constraints that guide our design choices. 
First, model memory capacity is bounded; arbitrarily increasing parameter counts would violate the hardware memory constraints of our inference servers. Second, training speed must be preserved. Any architectural change that significantly increases training time stalls the production pipeline and reduces overall engineering velocity. Third, serving latency must remain bounded, as increases in inference latency directly degrade the user experience.

The current production baseline relies on a shared Multi-Layer Perceptron (MLP) backbone. While operationally efficient, this architecture forces a single latent representation to satisfy disparate content distributions. This makes the model highly susceptible to negative transfer between conflicting objectives and minority-type collapse, where dominant content types drown out underrepresented ones. 
To break this bottleneck within our fixed operational budgets, we propose the \underline{H}eterogeneity-\underline{A}daptive Multi-gated \underline{M}ixture-\underline{o}f-\underline{E}xperts (\textbf{HA-MoE}). Rather than drastically expanding the model size or redesigning complex feature ingestion pipelines, HA-MoE focuses on improving the utilization of existing explicit heterogeneity signals ($h$) within the model backbone. 

\textbf{Model Design.} We adopt a dense, expert-based multi-task learning framework to predict the $K$ distinct user interaction tasks. Each expert is a feed-forward network (FFN). By utilizing a dense multi-gate MoE backbone~\cite{ma2018modeling}, we effectively decompose the monolithic shared representation into specialized experts while maintaining robust convergence and high training speed. To further adapt to the content heterogeneity without significant operational overhead, HA-MoE introduces a heterogeneity-aware adaptation mechanism within both the gating networks and the expert layers.

As illustrated in Figure~\ref{fig:ha_moe}, for a model consisting of $N$ shared experts $\{E_n\}_{n=1}^N$, the final predicted probability $p_t$ for task $t \in \mathcal{T}$ is produced by passing the gated sum of the modulated experts through a task-specific tower $T_t$:
\begin{equation}
p_t = T_t \left( \sum_{n=1}^N g_t(x, h)_n \cdot \tilde{E}_n(x, h) \right)
\end{equation}
where $g_t(x, h)_n$ represents the scalar gating probability assigned to expert $n$ for task $t$, and $\tilde{E}_n(x, h)$ is the heterogeneity-adapted output of the $n$-th expert. This formulation relies on two key components to leverage heterogeneity. First, \textbf{Heterogeneity-Aware Gating (HA-Gating)} explicitly incorporates $h$ into the gating input to guide expert utilization for each task $t$:
\begin{equation}
g_t(x, h) = \text{Softmax}(W_t \Phi(x, h) + b_t)
\end{equation}
where $W_t \in \mathbb{R}^{N \times d_\Phi}$ is the gating weight matrix, $b_t \in \mathbb{R}^N$ is the corresponding bias vector, and $\Phi(x, h) \in \mathbb{R}^{d_\Phi}$ represents a lightweight fusion function, integrating the dense features $x \in \mathbb{R}^d$ with the explicit heterogeneity signals $h \in \mathbb{R}^{d_h}$ into a joint representation of dimension $d_\Phi$. By structurally emphasizing the influence of $h$ within this representation, the gating mechanism actively steers expert utilization across diverse content segments. This encourages a robust alignment between specialized experts and heterogeneous interaction patterns, a dynamic that is vital for protecting minority segments from being overshadowed by dominant content segments.

Second, to decouple expert capacity from distribution-specific biases, we introduce \textbf{Heterogeneity-Driven Linear Modulation (HDLM)} to dynamically adapt the expert representations. Inspired by conditional affine transformations \cite{perez2018film}, HDLM computes a scale vector $\gamma_n(h)$ and a shift vector $\beta_n(h)$ dependent on the explicit heterogeneity signals:
\begin{equation}
\gamma_n(h) = W_{\gamma, n} h + b_{\gamma, n}, \quad \beta_n(h) = W_{\beta, n} h + b_{\beta, n}
\end{equation}
where $W_{\gamma, n}, W_{\beta, n} \in \mathbb{R}^{d_e \times d_h}$ are the modulation weight matrices, and $b_{\gamma, n}, b_{\beta, n} \in \mathbb{R}^{d_e}$ are the respective bias vectors for expert $n$. While this context-aware modulation can be applied at different stages, we apply it to the final expert layer to preserve representational expressiveness while maintaining training speed and stability in production. The adapted representation $\tilde{E}_n(x, h) \in \mathbb{R}^{d_e}$ is then computed via a late-stage affine transformation:
\begin{equation}
\tilde{E}_n(x, h) = \gamma_n(h) \odot E_n(x) + \beta_n(h)
\end{equation}
where $\odot$ denotes element-wise multiplication. By contextually recalibrating these representations, the network can implicitly ``switch contexts'' based on the nature of the input. For instance, when processing a video, the modulation layer could steer the focus toward visual consumption patterns. For content from a followed creator, it could shift the predictive emphasis away from the global popularity priors that heavily influence unfollowed organic ranking, orienting instead toward user-creator affinity. For newer formats like generative AI summary cards, the network could pivot its feature reliance, transitioning away from sparse historical interaction signals to focus on the rich semantics within the card.

\textbf{Model Optimization.} We use a combined training objective to calibrate absolute probabilities while explicitly preserving relative ordering:
\begin{equation}
\mathcal{L} = \alpha \mathcal{L}_{p} + (1 - \alpha) \mathcal{L}_{r}
\end{equation}
where $\alpha \in [0, 1]$ is a hyperparameter balancing the pointwise and pairwise losses. Given a training batch $\mathcal{B}$, the pointwise Binary Cross-Entropy (BCE) loss calibrates the $K$ prediction heads:
\begin{equation}
\mathcal{L}_{p} = \sum_{i \in \mathcal{B}} \sum_{t \in \mathcal{T}} w_t  \mathcal{L}_{\text{BCE}}(y_{t,i}, p_{t,i})
\end{equation}
where $w_t$ is the task weight, $y_{t,i}$ is the ground-truth label, and $p_{t,i}$ is the predicted probability for task $t$ on item $i$. To explicitly encourage the model to rank engaged items higher than non-engaged ones, we incorporate a pairwise RankNet loss~\cite{burges2005learning}:
\begin{equation}
\mathcal{L}_{r} = \frac{1}{|\mathcal{B}^+| \times |\mathcal{B}^-|} \sum_{i \in \mathcal{B}^+} \sum_{j \in \mathcal{B}^-} \log\left(1 + e^{-(s_i - s_j)}\right)
\end{equation}
where $\mathcal{B}^+$ and $\mathcal{B}^-$ denote the disjoint sets of items within the batch that received positive and negative feedback, respectively, and $s_i$ represents the model logit for item $i$. 
In production, we adapt this loss in several practical ways. Pairs are constructed per task from co-impressed items within the same user session, retaining only positive-negative pairs and applying additional filtering to remove noisy combinations. To further stabilize the gradients, the logit margins are scaled appropriately, and the final loss is dynamically normalized based on the volume of valid pairs in each batch.

\subsection{Continuous Production Monitoring via LENS}
\label{sec:lens_framework}
Within Discover, a persistent challenge in deploying unified multi-task architectures is their ``black box'' nature; a single shared representation obscures how the model simultaneously balances heterogeneous content distributions and conflicting task objectives. 

Adopting the HA-MoE framework unlocks a unique mechanism to ``unbox'' these internal dynamics for deeper interpretability. We introduce the \underline{L}atent \underline{E}xpert \underline{N}etwork \underline{S}pecialization (\textbf{LENS}) framework, which serves two valuable analytical functions: providing interpretable visual diagnostics of content-level specialization, and offering a quantitative analysis of task specialization strategies across different model states.

\textbf{Activation Slicing for Visualizing Specialization.} 
To examine how HA-MoE disentangles heterogeneous content, we extract the gating activations for each prediction task and slice them into separate content-by-expert matrices. Crucially, these activation probabilities are row-normalized per content type to quantify the proportional contribution of each expert, independent of absolute traffic. This yields a visualization of the model's internal division of labor, capturing both shared foundational representations for universal behaviors and distinct expert utilization patterns for specific content types (e.g., Web Articles, UGC, and Videos).

\textbf{Permutation-Invariant Matching for Specialization Analysis.} 
While visually diagnosing a single model state provides intuitive insights, managing continuous industrial pipelines benefits greatly from automated observability. As deployments increasingly value system predictability and stable internal representations \cite{fedus2022review, anelli2021perform}, scaling this analysis to compare multiple model snapshots provides critical diagnostic value in heterogeneous environments. While prior work addresses cross-run representation alignment through frameworks leveraging Multi-Level Optimal Transport \cite{shah2025representational} to map internal states, MoE architectures present a unique challenge. Because expert networks are permutation-invariant and highly sensitive to initializations, directly comparing parameter matrices across models is intractable.  

To automate the comparison of task specialization, we introduce \underline{P}ermutation-\underline{I}nvariant \underline{E}xpert \underline{M}atching (\textbf{PIEM}). Each expert $n$ is characterized by a behavioral profile vector $\mathbf{v}_n$, representing its relative contribution distributed across the $K$ prediction tasks. By normalizing these vectors into valid probability distributions, we quantify their functional difference using the Jensen-Shannon Divergence (JSD). To align the $N$ experts between a baseline model snapshot A and a comparison snapshot B, we employ the Hungarian algorithm~\cite{kuhn1955hungarian} to solve the 1-to-1 bipartite matching problem. This yields the optimal permutation $\sigma^*$ that minimizes the total JSD:
\begin{equation}
\sigma^* = \arg\min_{\sigma} \sum_{n=1}^{N} \text{JSD}(\mathbf{v}_{A,n} \parallel \mathbf{v}_{B, \sigma(n)})
\end{equation}
The final PIEM score is computed as the average similarity across these optimally matched expert pairs:
\begin{equation}
\text{PIEM}(A, B) = \frac{1}{N} \sum_{n=1}^{N} \left(1 - \sqrt{\text{JSD}(\mathbf{v}_{A,n} \parallel \mathbf{v}_{B, \sigma^*(n)})}\right)
\end{equation}

Crucially, PIEM functions as an automated diagnostic signal for monitoring functional heterogeneity, rather than a direct proxy for model performance. Given the highly non-convex optimization landscape of MoE architectures, multiple equivalent local optima exist. When comparing independent retrains, a model may converge to different parameterizations and expert allocation strategies while achieving comparable downstream evaluation metrics. Therefore, a low PIEM score between independent runs does not necessarily imply a degraded model; rather, it simply quantifies that the models have discovered divergent strategies. However, in cases of functional collapse, where a model fails to exhibit expected heterogeneity and lacks distinct task specialization, PIEM successfully captures this absence of functional differentiation, resulting in a distinctly low PIEM score against a healthy baseline.

In continuous, warm-start training pipelines, PIEM provides further diagnostic utility for tracking shifts in expert utilization over time. When evaluated over a temporal sliding window, PIEM quantifies how expert specialization adapts to gradual data drift. In this setting, the optimal permutation $\sigma$ is theoretically expected to remain the identity mapping ($\sigma(n) = n$). Significant deviations can flag potentially problematic internal model behavior, though attributing them to model issues versus rare, substantial data drift remains a known limitation. By operating entirely label-free on lightweight behavioral profile vectors, PIEM offers machine learning engineers an early quantitative signal to monitor internal model heterogeneity well before full offline evaluations complete.
\section{Experiments}

\subsection{Experimental Settings}
\subsubsection{Evaluation Metric: DL-AUC}
\label{sec:evaluation_framework}

Building upon a stable training foundation, evaluating ranking performance in modern recommender systems necessitates a multi-faceted approach. Alongside online A/B testing, offline assessment has historically relied on ROC AUC~\cite{fawcett2006introduction} and NDCG~\cite{jarvelin2002cumulated}. However, the increasing importance of distributionally-informed evaluation~\cite{ekstrand2024distributionally} and provider fairness~\cite{boratto2021interplay, wu2025preserving} has driven the adoption of metrics like Cross-AUC (xAUC)~\cite{kallus2019fairness}. By quantifying disparities in bipartite ranking across distinct groups, xAUC provides a vital foundation for explicitly assessing cross-type ranking correctness in highly heterogeneous feeds. In our evaluation, we use ``cross-segment'' and ``cross-type'' interchangeably, with segments defined by content type.

In industrial systems, standard global metrics often mask severe structural failures. For example, a model might trivially increase its overall AUC by uniformly inflating the scores of a data-rich majority segment (e.g., web articles) while burying high-quality minority content (e.g., videos). To provide a single, automated decision signal for model selection, we introduce \textbf{Dual-Level AUC (DL-AUC)}, which measures performance at two levels by balancing overall ranking performance and cross-segment correctness.

Ideally, such an objective would compute the exact Group AUC (GAUC) to ensure rigorous personalized ranking per user, alongside a traffic-weighted sum of xAUC to preserve exact margins. However, calculating exact GAUC across massive user-item subgroups is computationally prohibitive for fast-iteration industrial pipelines. Therefore, we adopt a highly scalable approximation for production:
\begin{equation}
\text{DL-AUC} = \lambda \cdot \text{Micro-AUC} + (1 - \lambda) \cdot \text{Macro-xAUC},
\end{equation}
where $\lambda$ denotes a business policy weight controlling the trade-off between overall utility and ecosystem health. To formalize these components, let $\mathcal{D}^+$ and $\mathcal{D}^-$ denote the global sets of positive and negative interactions in the evaluation data, and $p_i, p_j$ be the predicted probabilities for items $i$ and $j$. Let $\mathbb{I}(\cdot)$ be the indicator function. The components are defined as follows:

\begin{itemize}
    \item \textit{Micro-AUC}: Reflects global ranking quality under the empirical traffic distribution. It measures the ability to accurately rank engaging items across the entire feed:
    \begin{equation}
    \text{Micro-AUC} = \frac{1}{|\mathcal{D}^+| |\mathcal{D}^-|} \sum_{i \in \mathcal{D}^+} \sum_{j \in \mathcal{D}^-} \mathbb{I}(p_i > p_j)
    \end{equation}

    \item \textit{Macro-xAUC}: The unweighted average xAUC across all pairs of segments. In our use case, we specifically evaluate all pairs of key content types. Let $\mathcal{C}$ be the set of these distinct content types. For any two types $A, B \in \mathcal{C}$ where $A \neq B$, $\text{xAUC}(A, B)$ evaluates the probability that a positive item from $A$ outranks a negative item from $B$. As a structural metric, it prevents majority class domination:
    \begin{equation}
    \text{xAUC}(A, B) = \frac{1}{|\mathcal{D}_A^+| |\mathcal{D}_B^-|} \sum_{i \in \mathcal{D}_A^+} \sum_{j \in \mathcal{D}_B^-} \mathbb{I}(p_i > p_j)
    \end{equation}
    \begin{equation}
    \text{Macro-xAUC} = \frac{1}{|\mathcal{C}|(|\mathcal{C}| - 1)} \sum_{A, B \in \mathcal{C}, A \neq B} \text{xAUC}(A, B)
    \end{equation}

\end{itemize}

By blending these signals into a single scalar, DL-AUC enables mathematically grounded deployment decisions. To illustrate why this formulation is critical for automated guardrails, we highlight how DL-AUC effectively navigates three common industrial scenarios (using web articles and videos as an intuitive running example):

\begin{itemize}
    \item \textit{Dominant Score Inflation}: If a model artificially inflates the scores of a majority type to maximize global AUC, the unweighted Macro-xAUC drops significantly. Because AUC is bounded at $1.0$, the marginal cross-type gains of the inflated type ($\text{Web}^+$ vs.\ $\text{Vid}^-$) cannot mathematically offset the severe degradation of the suppressed type ($\text{Vid}^+$ vs.\ $\text{Web}^-$). DL-AUC automatically penalizes this asymmetry, preventing the launch of models that harm ecosystem health.
    \item \textit{Clean Global Improvement}: If a model improves underlying representations for both content types, both intra-type ranking and cross-type margins increase. DL-AUC rises significantly, clearly quantifying the true model gain.
    \item \textit{Within-Type Gain}: If a specific content type's ranking improves internally without altering cross-type margins, Macro-xAUC remains stable while Micro-AUC rises. DL-AUC correctly rewards this as a safe utility gain.
\end{itemize}

\subsubsection{Implementation Details}

Deploying the HA-MoE architecture in an industrial production system requires balancing model capacity against strict operational constraints: model storage is constrained, training speed must be preserved, and serving latency must remain bounded. We adopt a dense MMoE rather than a sparse one~\cite{fedus2022review} to balance model performance against computational overhead. Each expert is a multi-layer feed-forward network (FFN) with ReLU activations. The optimal number of experts ($N=4$, selected from $\{2, 4, 8, 16\}$) and other architectural hyperparameters were determined via grid search to maximize representational power within these fixed budgets. For our initial production deployment, the fusion function $\Phi(x, h)$ within the gating network is implemented as a direct concatenation $[x \mathbin\Vert h]$. This lightweight design mitigates optimization instability and supports stable, efficient convergence. While we are actively exploring more expressive gating mechanisms, this low-complexity baseline, designed to optimize the trade-off between model performance and training stability, effectively drives the substantial empirical gains reported. To encourage diverse expert utilization and mitigate representation collapse, we apply orthogonal initialization to the expert networks.

This architecture is trained on a massive, proprietary dataset of Discover user interaction logs, optimized end-to-end with an Adam-family optimizer. The training corpus spans an extended temporal window, effectively capturing both short-term trends and mid-to-long-term user-item interaction patterns. Additionally, the training logs undergo positive upsampling and negative downsampling to mitigate inherent class imbalance.

To ensure fair comparisons, all optimization hyperparameters by default inherit pre-established production configurations and remain identical across all models. This includes the multi-task loss balancing coefficient $\alpha$ and the initial values of the 11 task weights $w_t$ before they are dynamically adjusted during training via gradient norms~\cite{chen2018gradnorm}. Finally, the policy parameter $\lambda$ in the DL-AUC computation is configured to $0.8$, assigning global utility $4\times$ the weight of cross-type ranking correctness. This reflects an operational policy of prioritizing overall ranking utility, with cross-type correctness serving as a safety floor; other contexts may tune $\lambda$ differently.

\textbf{Deployment Overhead.} Relative to the Shared MLP production baseline, HA-MoE increases model size by under 5\%, maintains comparable training speed within measurement noise, and adds under 0.5\% to end-to-end serving latency at both p50 and p95. As end-to-end latency reflects the full serving pipeline, these results confirm that HA-MoE fits comfortably within fixed production budgets.

\subsection{Offline Evaluation}


We evaluated HA-MoE on a sampled 7-day holdout dataset comprising approximately 10 million global examples. 
We group the model outputs of our 11 distinct prediction tasks into two macro-categories: \textbf{pInterest} (representing positive engagement predictions, such as clicks and likes) and \textbf{pDisinterest} (representing negative engagement predictions, such as dismissals and blocks). Table \ref{tab:offline_results} reports the aggregated DL-AUC for the tasks within each category.


\begin{table}[!t]
\caption{Offline performance comparison, measured by Aggregated DL-AUC.}
\label{tab:offline_results}
\begin{tabular}{lcc}
\toprule
\textbf{Model} & \textbf{pInterest} & \textbf{pDisinterest} \\
\midrule
Shared MLP & 0.679 & 0.939 \\
Standard MMoE & 0.686 & 0.934 \\
HA-MoE w/o HA-Gating & 0.689 & \textbf{0.949} \\
HA-MoE w/o HDLM & 0.690 & 0.945 \\
\textbf{HA-MoE} & \textbf{0.691} & \textbf{0.949} \\
\bottomrule
\end{tabular}
\end{table}

The results in Table~\ref{tab:offline_results} highlight a fundamental multi-task optimization conflict inherent in the baseline architectures. While the Standard MMoE improves positive engagement predictions over the Shared MLP, it regresses on negative predictions. In contrast, HA-MoE effectively breaks this trade-off. The ablation results further show that both heterogeneity-aware gating and the HDLM layer contribute to these gains, with the full model achieving the best aggregate performance across both task categories. By explicitly injecting heterogeneity context through both components, our proposed architecture secures the positive engagement gains while simultaneously reversing the regression in negative predictions. The pDisinterest DL-AUC score reaches 0.949, while HA-MoE achieves the best overall performance across diverse tasks without succumbing to negative transfer.

We also examine a specific historical challenge: the systemic underranking of videos compared to web articles. Using \texttt{pDisinterest} as an example, we demonstrate how global AUC masks severe cross-type biases because it is heavily dominated by high-volume, intra-type comparisons. Table \ref{tab:xauc_deepdive} provides a direct xAUC comparison of this cross-type ranking correctness to reveal the performance gap. For brevity within the table, Web Articles and Videos are abbreviated as Web and Vid, respectively, and HA-MoE denotes the complete HA-Gating + HDLM architecture.

\begin{table}[!t]
\caption{Cross-type xAUC on pDisinterest tasks: Web Articles (Web) vs. Videos (Vid).}
\label{tab:xauc_deepdive}
\begin{tabular}{lccc}
\toprule
\textbf{Model} & \textbf{Web$^+$ vs. Vid$^-$} & \textbf{Vid$^+$ vs. Web$^-$} & \textbf{$\Delta$ Gap ($\downarrow$)} \\
\midrule
Shared MLP & 0.971 & 0.830 & 0.141 \\
\textbf{HA-MoE} & 0.960 & 0.900 & \textbf{0.060} \\
\bottomrule
\end{tabular}
\end{table}
In the Shared MLP baseline, poor-quality web articles frequently outranked high-quality videos, resulting in a significant 0.141 gap between opposing cross-type pairs. HA-MoE corrects this imbalance, boosting the minority pair (Vid$^+$ vs. Web$^-$) to 0.900 and narrowing the gap to 0.060. This confirms that HA-MoE prevents bad web articles from unfairly suppressing good videos, driving the aggregate DL-AUC improvements that are missed by standard global AUC.

This quantitative analysis offers insight into large-scale heterogeneous recommender systems: explicitly modeling heterogeneity within both the experts and the gating network provides the representational capacity needed to disentangle competing ranking patterns across diverse content types.

\subsection{Interpretability and System Observability}
Beyond offline ranking performance, deploying unified multi-task architectures across heterogeneous environments benefits greatly from ``unboxing'' internal dynamics to understand the model's division of labor and track its evolution across continuous retraining. We demonstrated these capabilities using our proposed LENS framework.

\begin{figure}[!t]
\centering
\includegraphics[width=\linewidth]{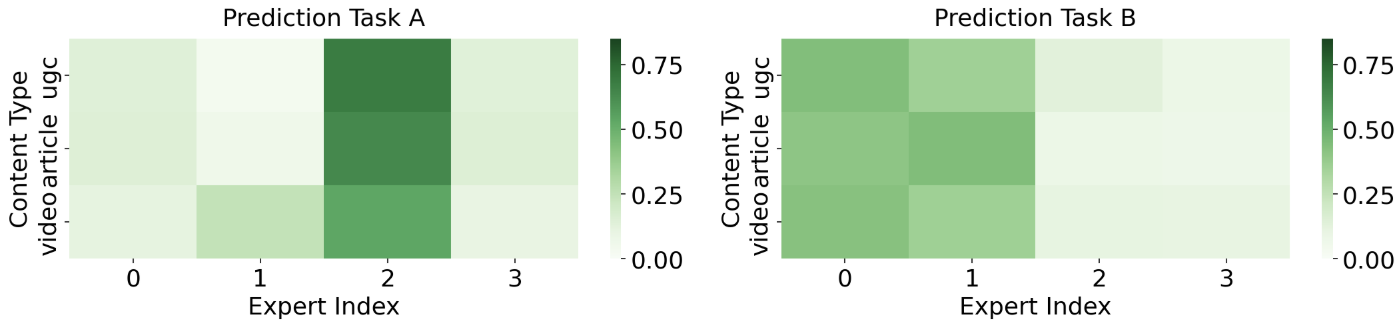}
\Description{Two heatmaps showing universal task activation. Task A: general positive preference. Task B: sparse survey feedback.}
\caption{Expert Activation for Universal Tasks. Task A and Task B represent dense and sparse universal actions, respectively. Universal user actions tend to utilize shared experts across all content types.}
\label{fig:lens_slicing_1}
\end{figure}

\begin{figure}[!t]
\centering
\includegraphics[width=\linewidth]{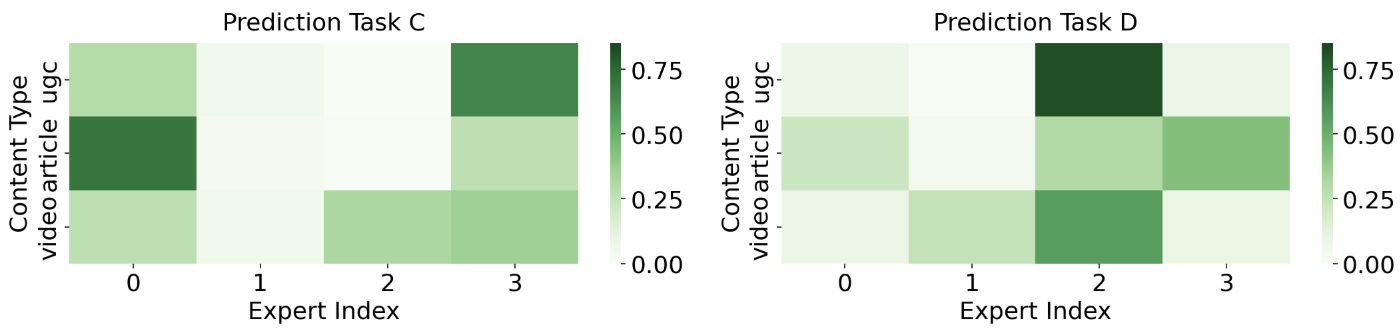}
\Description{Two heatmaps showing specialized task activation, demonstrating divergent activation across content types for Task C and Task D.}
\caption{Expert Activation for Specialized Tasks. Tasks C and D represent format-dependent interactions, demonstrating specialized, divergent expert utilization for different content types.}
\label{fig:lens_slicing_2}
\end{figure}

\textbf{Unboxing Internal Dynamics via Activation Slicing.} 
To understand how HA-MoE disentangles heterogeneous content, we utilized LENS's activation slicing, sampling three key content types for illustrative clarity. As shown in Figures \ref{fig:lens_slicing_1} and \ref{fig:lens_slicing_2}, this visualization reveals distinct expert utilization strategies that vary significantly depending on the nature of the prediction task. We observe three primary patterns:

\begin{itemize}
    \item \textit{Universal Prediction Tasks}: For broadly applicable, format-agnostic user actions, such as expressing a ``like'' via a heart button available directly on the main feed for every item (Task A), the model generally demonstrates strong common learning. As shown in Figure \ref{fig:lens_slicing_1}, a single expert predominantly handles this prediction task across all content types.
    \item \textit{Universal but Sparse Tasks}: For tasks that apply to all content but have much sparser positive labels, such as explicit positive survey feedback (Task B), the model allocates two shared experts. This suggests that while the task spans all content types, the model leverages slightly more specialized capacity to capture distinct sub-patterns within the sparse feedback.
    \item \textit{Specialized/Format-Dependent Tasks}: For behaviors inherently tied to specific content types, such as clicking into a detailed view (Task C) or expanding a text block (Task D), the heatmaps display highly specialized activation patterns. As illustrated in Figure \ref{fig:lens_slicing_2}, different experts tend to activate predominantly for different content types.
\end{itemize}

\begin{figure}[!t]
\centering
\includegraphics[width=\linewidth]{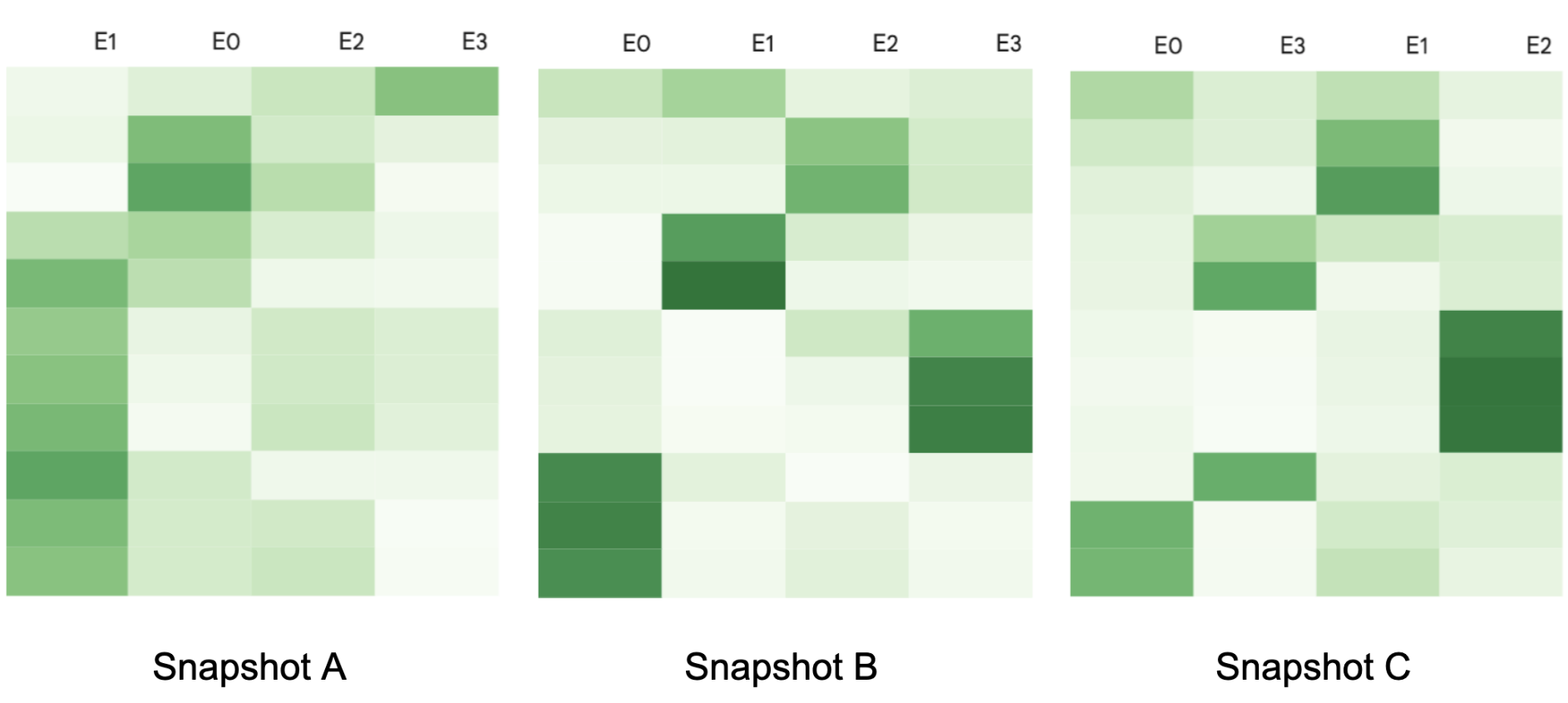}
\Description{Three task-by-expert activation heatmaps for model snapshots A, B, and C. Rows are prediction tasks and columns are experts; darker cells indicate stronger activation. Snapshots B and C display distinct, distributed activation patterns across multiple experts, while snapshot A collapses most tasks onto a dominant expert, illustrating how PIEM distinguishes healthy specialization from functional collapse.}
\caption{Task specialization tracking using task-expert activation profiles for three model snapshots. Snapshots B and C exhibit functional, distributed expert roles, yielding a high PIEM score when matched. Snapshot A suffers from activation imbalance, resulting in a low PIEM score when compared to a healthy baseline.}
\label{fig:lens_matching}
\end{figure}

\textbf{Tracking Specialization via Permutation-Invariant Matching.}
Beyond initially learning specialized expert roles, robust system observability requires tracking how these functional assignments persist or adapt across continuous production cycles. To monitor expert task specialization, we utilize PIEM as a lightweight diagnostic. Figure \ref{fig:lens_matching} illustrates task-level expert utilization profiles for three model snapshots (Snapshots A, B, and C) to demonstrate how the metric reflects internal structural shifts.

The heatmaps display the task-level expert utilization profiles for each model snapshot, where rows denote prediction tasks and columns denote experts. When comparing model snapshots B and C, we observe that both snapshots exhibit distinct task specialization, yielding a high $\text{PIEM}(B, C)$ score. This indicates stable structural behavior and the maintenance of specialized expert roles. Conversely, the comparison between snapshot A and snapshot B yields a low $\text{PIEM}(A, B)$ score. Examining model snapshot A's profile reveals a notable lack of expected functional heterogeneity, as the network fails to develop differentiated expert roles. This absence of functional differentiation---where tasks remain largely uniform and often default to a single dominant expert---impedes the alignment of valid functional pairs under the optimal permutation $\sigma^*$, consequently reducing the score.
This structural observation aligns with our empirical results, as the initial, unoptimized configuration of snapshot A yielded lower evaluation metrics than B and C.

In practice, this diagnostic metric provides an interpretable signal that helps engineers identify unexpected shifts in expert specialization and investigate potential instability when needed. Encompassing both Activation Slicing and Permutation-Invariant Matching, the LENS framework bridges the gap between raw model weights and system observability, supporting a highly interpretable and operationally transparent architecture.

\subsection{Online A/B Testing}
With offline evaluation criteria satisfied and production readiness established, the HA-MoE architecture was deployed for online A/B testing to assess its real-world impact. We conducted multiple 7-day experiments on 1\% of live traffic within Discover, a recommender platform serving hundreds of millions of daily active users and billions of monthly active users globally. These tests evaluated the proposed architecture against the existing production baseline, a shared MLP-based multi-task prediction model. All experiments were monitored under standard production guardrails.

\begin{table}[!t]
\caption{Online A/B test results, reported as lift over production $\pm$ 95\% CI half-width.}
\label{tab:online_results}
\begin{tabular}{lc}
\toprule
\textbf{Feed Activity Metrics} & \textbf{Lift ($\pm$ CI)} \\
\midrule
DAU & +0.22\% $\pm$ 0.11\% \\
Viewed Impressions & +0.48\% $\pm$ 0.34\% \\
\midrule
\textbf{Feed Exploration Metrics} & \textbf{Lift ($\pm$ CI)} \\
\midrule
Scroll Depth & +0.34\% $\pm$ 0.25\% \\
Diverse Feed Rate & +0.36\% $\pm$ 0.03\% \\
Diverse Engagement Rate & +0.54\% $\pm$ 0.07\% \\
\bottomrule
\end{tabular}
\end{table}

\textbf{Production Impact:} The online results in Table~\ref{tab:online_results} demonstrate that lifts in feed activity metrics closely align with the offline DL-AUC improvements. By achieving more accurate predictions for both positive and negative user interactions across different content types, the model drove higher user engagement, as reflected in the positive lifts for Daily Active Users (DAU) and Viewed Impressions. These improvements were consistent across multiple live experiments, with no observed regressions in other core operational metrics. Beyond basic engagement, we observed notable gains in feed exploration metrics, specifically average Scroll Depth, Diverse Feed Rate, and Diverse Engagement Rate. Together, these metrics indicate that users navigated deeper into their feeds, were served a more diverse set of content, and interacted with a broader variety of content. Furthermore, we observed that the suppression of certain historically under-ranked traffic was effectively mitigated. These gains validate that improved heterogeneous ranking translates to tangible benefits for content discovery and overall ecosystem health. Together, these results confirm the practical value of the proposed approach for industrial-scale heterogeneous ranking, supporting its evolution from an experimental prototype to a production-ready architecture.
\section{Conclusion and Future Work}
This paper addresses the challenge of heterogeneous ranking in industrial-scale recommender systems that unify diverse content from the open web. We present an end-to-end production case study from Google Discover, detailing the design, deployment, and evaluation of a unified multi-task ranking system for heterogeneous feeds.

Our contributions span key stages of the model lifecycle. At the model level, we introduce HA-MoE, a heterogeneity-adaptive multi-gated mixture-of-experts architecture that integrates explicit heterogeneity context into both the gating networks and expert representations, enabling effective specialization without exceeding strict production constraints. To ensure operational reliability, we propose LENS, an observability framework that ``unboxes'' the model's internal division of labor and quantitatively tracks functional task specialization over time. Finally, we establish DL-AUC, a heterogeneity-aware evaluation metric that captures cross-type ranking correctness alongside overall ranking performance.

Both offline evaluation and online A/B experiments demonstrate consistent improvements, translating to lifts in feed activity and exploration metrics. Notably, these improvements are achieved without regressions in operational costs, validating a practical design pattern for industrial-scale heterogeneous recommender systems and underscoring the value of explicitly modeling heterogeneity in unified ranking.

While this work establishes a strong foundation, managing heterogeneity at industrial scale remains an evolving challenge. Future work will focus on advancing expressive gating, exploring scalable sparse routing, and refining expert modulation mechanisms to further enhance model representational power. These efforts will also deepen our understanding of the interplay between shared learning and heterogeneity-driven specialization. We aim to extend heterogeneity modeling beyond content-centric signals to incorporate rich user-level context and broader business objectives. Finally, we plan to explore listwise optimization and LLM-enhanced approaches to advance unified heterogeneous ranking from item-level decisions to feed-level orchestration, capturing intra-feed dependencies and engagement patterns across user sessions.

\begin{acks}
The authors acknowledge the use of Google Gemini 3 to refine the manuscript's text. All AI-assisted revisions were manually reviewed by the authors, who assume full responsibility for the final work.


\end{acks}

\clearpage
\bibliographystyle{ACM-Reference-Format}
\bibliography{hag-moe-references}

\clearpage

\end{document}